\documentclass[twocolumn]{article}
\usepackage[table]{xcolor}
\usepackage{graphicx}
\usepackage{amssymb}
\usepackage[numbers]{natbib}
\usepackage{authblk}
\usepackage{textgreek}
\usepackage[utf8]{inputenc}
\usepackage[english]{babel}

\usepackage{hyperref}
\usepackage{booktabs}
\hypersetup{
    unicode,
    colorlinks=true,
    linkcolor=linkcolor,
    citecolor=linkcolor,
    filecolor=linkcolor,
    urlcolor=linkcolor,
}
\usepackage{color,colortbl}
\definecolor{linkcolor}{rgb}{0.0,0.3,0.5}
\DeclareGraphicsExtensions{.bmp,.png,.jpg,.pdf}
\usepackage{verbatim}
\usepackage[normalem]{ulem}
\usepackage{orcidlink}
\usepackage{booktabs}
\usepackage{soul}
\usepackage{enumitem}

\graphicspath{{./figs/}}

\usepackage{newtxtext,newtxmath}
\usepackage[T1]{fontenc}
\usepackage{amsmath, amssymb, color, units}
\usepackage[table]{xcolor}
\usepackage{colortbl}
\usepackage{etoolbox}

\usepackage{etoolbox}

\newcommand{\UCSD}{Department of Astronomy and Astrophysics, University of California, San Diego, La Jolla, CA 92093, USA}
\newcommand{\CfA}{Center for Astrophysics \textbar{} Harvard \& Smithsonian, 60 Garden St., Cambridge, MA 02138, USA}

\newcommand{\NCSA}{National Center for Supercomputing Applications, University of Illinois Urbana-Champaign, 1205 W. Clark St, Urbana, IL 61801, USA}
\newcommand{\USRSE}{The US Research Software Engineer Association, 1000 Broadway, Suite \#480, Oakland, CA 94607}
\newcommand{\UofT}{David A. Dunlap Department of Astronomy \& Astrophysics, University of Toronto, Toronto, ON M5S 3H4, Canada}
\newcommand{\DI}{Dunlap Institute for Astronomy \& Astrophysics, University of Toronto, Toronto, ON M5S 3H4, Canada}
\newcommand{\FAIR}{FAIR Data Innovations Hub, California Medical Innovations Institute, San Diego, CA, 92121 USA}
\newcommand{\UCBerk}{University of California Berkeley, Berkeley, CA 94720, USA}
\newcommand{\CCA}{Center for Computational Astrophysics, Flatiron Institute, New York, NY 10010, USA}
\newcommand{\ICER}{Institute for Cyber-Enabled Research, Michigan State University,
East Lansing, Michigan 48824, USA}
\newcommand{\SDSC}{San Diego Supercomputer Center, University of California, San Diego, La Jolla, CA 92093, USA}
\newcommand{\UIUC}{University of Illinois Urbana-Champaign, Urbana, IL, USA}
\newcommand{\ASCL}{Astrophysics Source Code Library | University of Maryland College Park, USA}
\newcommand{\Heidelberg}{Universit\"at Heidelberg, Zentrum f\"ur Astronomie (ZAH), Institut f\"ur Theoretische Astrophysik, Albert Ueberle Str. 2, 69120, Heidelberg, Germany}
\newcommand{\SciOS}{SciOS}

\begin{document}

%%
%% ─────────────────────────────────────────────────────────────────────────────
%%  FRONT MATTER
%% ─────────────────────────────────────────────────────────────────────────────
%%

\title{Best practices in software citation}

% Authors — fill in as the author list is confirmed
\author[1,2,3]{Phil R. Van-Lane\orcidlink{0009-0009-4567-9946}\thanks{pvanlane@ucsd.edu}}
\author[1]{Floor S. Broekgaarden\orcidlink{0000-0002-4421-4962}}

\author[4]{Daniel S. Katz\orcidlink{0000-0001-5934-7525}}
\author[5]{Bhavesh Patel\orcidlink{0000-0002-0307-262X}}
\author[6]{Pengyin Shan\orcidlink{0009-0009-6309-380X}}
\author[7]{Jonathan Starr}
\author[8]{Samantha Teplitzky\orcidlink{0000-0001-7071-332X}}
\author[9]{Peter K.G. Williams\orcidlink{0000-0003-3734-3587}}

\author[10]{Alice Allen\orcidlink{0000-0003-3477-2845}}
\author[11]{Lucas M. de S\'{a}\orcidlink{0000-0003-3109-9042}}
\author[12]{Andrew Fullard\orcidlink{0000-0001-7343-1678}}
\author[13,14]{Sandra Gesing\orcidlink{0000-0002-6051-0673}}
\author[15]{Tom Wagg\orcidlink{0000-0001-6147-5761}}
\author[14,1]{Andrea Zonca\orcidlink{0000-0001-6841-1058}}

\affil[1]{\UCSD}
\affil[2]{\UofT}
\affil[3]{\DI}

\affil[4]{\UIUC}
\affil[5]{\FAIR}
\affil[6]{\NCSA}
\affil[7]{\SciOS}
\affil[8]{\UCBerk}
\affil[9]{\CfA}
\affil[10]{\ASCL}
\affil[11]{\Heidelberg}
\affil[12]{\ICER}
\affil[13]{\USRSE}
\affil[14]{\SDSC}
\affil[15]{\CCA}

\maketitle

\begin{abstract}
Software is both a foundational tool and a primary output of modern computational research, yet citation practices for software remain inconsistent, incomplete, and rarely machine-actionable. 
Existing infrastructure designed for paper and data citation does not adequately serve the distinct needs of software citation, leaving a gap that impedes reproducibility, misattributes scholarly credit, and obscures the labor embedded in research pipelines. 
Drawing on a NASA-funded community workshop held in April 2026, we present an analysis of four interconnected themes: (I)~the cultural barriers to consistent citation practice; (II)~the need for clearer community norms and conventions; (III)~gaps in existing technical infrastructure and workflow; and (IV)~the emerging challenges posed by AI-assisted research. 
For each theme we identify targeted interventions and assign responsibility across stakeholder groups.
We conclude that meaningful progress requires simultaneous action on technical and cultural fronts. Journal editors and publishers represent the single highest-leverage point for accelerating this change, and correct citation must become the path of least resistance within researchers' existing workflows.
\end{abstract}

%%
%% ─────────────────────────────────────────────────────────────────────────────
\section{Introduction}
\label{sec:intro}
%% ─────────────────────────────────────────────────────────────────────────────
%%

Modern research depends on software at every stage of the
research process: from data reduction and numerical simulation to
statistical inference and visualization. Software tools represent years of
intellectual labor, often produced by teams of (early-career) researchers whose
professional futures depend on recognition of that work. Yet software citation
in the published literature is chronically inconsistent; software is often only mentioned by name in text. When formal citations do
appear, they frequently fail to meet the technical standards that would make
them machine-actionable, meaning the credit they convey cannot be
automatically aggregated, tracked, or attributed to the specific individuals
who built the code \citep[e.g.][]{Schindler2023}.

This situation persists not because researchers are indifferent to credit,
but because the infrastructure and norms designed for paper and data citation
have never been adequately extended to software \citep[e.g.][]{Smith2016}. Software
is not a static artifact like a journal article: it evolves across versions,
accumulates contributors over time, and sits within complex dependency
relationships that have no direct analogue in traditional publishing \citep{Katz2016software}. Meeting
the specific needs of software citation therefore requires both \textit{technical
innovation}---new tools, formats, and workflows---and \textit{cultural
change}---new expectations, enforcement mechanisms, and incentive structures.

A NASA funded software citation workshop (\url{https://softwarecitationworkshop.ucsd.edu}) held in April 2026 convened researchers, software developers,
librarians, scholarly communication professionals, and journal editors to
address this challenge. The workshop had two primary goals: (1)~to develop
improved community guidelines on how to cite software and how to make
software citable; and (2)~to identify and prioritize enhancements to the
\href{https://www.tomwagg.com/software-citation-station/}{Software Citation
Station} (SCS), the primary community tool for astronomy software citation.
Participants engaged in structured marketplace discussions across fourteen
topic areas (Appendix~A), and several working groups continued
tool development after the workshop concluded.

This paper synthesizes the workshop's findings and recommendations. The
overarching thesis is that \textit{software citation is necessary and its
needs are not served by existing infrastructure for paper and data citation;
the improvements proposed here balance technical and cultural change.} The workshop was not convened to address AI, and its recommendations do not depend on any particular trajectory for AI-assisted research. AI-assisted research nonetheless emerged as one of the themes in its own right, because it is already changing how software is cited and by whom, and the community will need to track these changes as software citation practice and infrastructure evolve. Sections~2
through 5 develop our four core themes. Section~6 identifies open questions and priority directions
for future work. We also describe the tools and resources produced by the workshop in the Appendix.

%%
%% ─────────────────────────────────────────────────────────────────────────────
\section{Theme I: Catalyzing Culture Change}
\label{sec:culture}
%% ─────────────────────────────────────────────────────────────────────────────
%%

Software citation serves several distinct but mutually reinforcing functions \citep[e.g.,][]{Smith2016,Katz2020}.
First, it enables \textit{attribution}, so that the
researchers who built the tools gain measurable, citable evidence of scholarly
impact, which can translate directly into funding, hiring, and career
recognition. Second, building on attribution, it supports \textit{sustainability}, as  developers who can demonstrate usage and
impact are better positioned to secure the resources needed to maintain and
extend their software. Third, it enables \textit{discoverability}, as machine-actionable citations
can be used to create a graph linking papers to the software they depend on, enabling new
forms of scholarly navigation, impact assessment, and software reuse
\citep{Niemeyer2016, Katz2018}.
Finally, it supports \textit{reproducibility}, so that a reader will know exactly which software, at which
version, was used in an analysis, which will help them (in principle) reconstruct the
computational environment and verify the results, though additional information is also needed beyond just the software itself, such as how it was used.

The question may be asked: why not just cite a related manuscript, be it a paper describing the software, a method paper, or a publication that motivated development of the software? This may seem reasonable since manuscript citations are thoroughly tracked by established platforms (Web of Science, Google Scholar, etc.) and plugged into common career metrics like the h-index. However, relying solely on a manuscript citation leaves gaps: many people contribute to software over time, and citing only a paper, rather than the version of the software used, could leave out credit for those who contributed after the paper was published, and furthermore, just citing a related manuscript is not as supportive of reproducibility as citing the specific version of the software itself. Direct software citation, alongside a potential paper citation, is therefore critical.

Despite its benefits, formal software citations, those appearing in
reference lists with persistent identifiers, remain rare relative to actual
software usage \citep{Howison2016seeing, Du2021softcite, Druskat:2024}. When software is mentioned at all, it is often acknowledged
informally in prose, with a bare URL or an informal reference that no indexing
system can automatically process. 
The scientific credit intended by the
authors of cited software frequently fails to reach those authors in any
trackable or attributable form \citep{Schindler2021, Schindler2023, Katz2020, Bouquin...2020ApJS..249....8B}.

%%
%% ─────────────────────────────────────────────────────────────────────────────
%% ─────────────────────────────────────────────────────────────────────────────
%%

\subsection{The Problem with the Status Quo}

The status quo will not change
dramatically without deliberate, coordinated interventions. Improvements in
tooling alone are insufficient: even when a clear workflow to correct citation
practice exists, researchers who do not know about it, do not prioritize it,
or who face friction at submission time will default to informal
acknowledgments or non-persistent URLs. Cultural change requires interventions
calibrated to the specific interests and constraints of each stakeholder group.

At present, researchers who \textit{try} to cite
software frequently fail to do so in ways that conform to technical standards \citep[e.g.][]{Howison2016seeing,  Schindler2023,  Bouquin...2020ApJS..249....8B}.
The citations they produce are often incomplete (missing version numbers or
DOIs), incorrectly formatted (prose acknowledgments rather than formal
reference-list entries), or misdirected (citing a paper \textit{about} the
software rather than the software itself). 
Meanwhile, most software mentions carry no formal citation. A manual analysis of 90 biology articles found that only 31-43\% of software mentions involved a formal citation \citep{Howison2016seeing}, and larger studies report rates as low as 16-25\%  \citep{Schindler2021, Schindler2023}. Even when a formal citation is present, it points to an article rather than the software itself in 84–89\% of cases \citep{Howison2016seeing, Schindler2023}. Because these studies count only software that is mentioned, they likely overestimate citation rates relative to actual use.
This gap between intention and outcome reflects both a knowledge deficit and a tooling
deficit.

Some publishers do encourage software citation, though uptake is difficult to monitor and measure, as most have focused more on the sharing of new software developed in the research than the citation of preexisting software used in it. For example, 66\% of submitted manuscripts to \textit{Nature Computational Science} from November 2023 to March 2024 reported having developed new code and 39\% wanted to share their code \citep{Pastrana2024}. Additionally, PLOS's Open Science Indicators dashboard reports that for 2025, 14.8\% of \textit{PLOS} papers shared code online, another 2.5\% of papers shared code through supplementary materials, and another 30.2\% of paper generated code but did not share it \citep{PLOS-OSI}. And finally, for papers published in \textit{Science} between 2021 and 2024, 23\% shared code, while 46\% did not generate code \citep{Science-software}. As an example from a more general perspective, the French Open Science Monitor analyzed papers published by French authors, and found that in 2024, 20\% mentioned sharing of code or software, and 49\% mentioned the use of code or software \citep{FrenchOSM}.

\subsection{The Stakeholder Landscape}

There are four principal categories of stakeholders, each with
distinct motivations and different capacities to influence software citation best practices.

\textit{Researchers} are the primary generators and consumers of software
citations. There are three distinct subgroups among those who use
software: those who would like to
adopt correct software citation practices if doing so was sufficiently clear and
frictionless, those who already attempt to cite software but do so inconsistently or
incorrectly due to a lack of knowledge or convenient tooling, and those who do not currently prioritize software citation and
would require stronger incentives or institutional mandates to change behavior.
Reaching all three groups requires fundamentally different approaches.

\textit{Software developers} have a direct stake in being cited but often
lack the knowledge or bandwidth to make their code easy to cite, as they are frequently in a junior role. Even when
developers provide citation instructions, the format can be inconvenient for users to follow or may fail to conform to the technical
standards that make citations machine-actionable. It is worth noting that developers are often researchers, so clearly differentiating the requirements for these two aspects is important to avoid confusion.

\textit{Journal editors and publishers} possess the most
leverage for change in the software citation ecosystem. Their ability to set and enforce submission
requirements means they can shift citation norms faster than any other actor.
Yet many editors lack the technical knowledge, resources, or institutional
support to enforce software citation standards rigorously, and many journals
have not yet adopted explicit software citation policies.

\textit{Funders and institutions} can incentivize behavior through grant
requirements and evaluation criteria. The extent to which software impact is
recognized in tenure and promotion processes, and in grant assessments, shapes
researchers' motivation to invest in citable software development. Currently
this recognition is minimal in most institutions.

\subsection{Stakeholder-Specific Interventions}

For \textit{researchers}, the primary interventions are educational and
tooling-based. Guidelines must be specific, short, and actionable for a practitioner finalizing a manuscript under deadline pressure. A practical
seven-step Software Citation Primer (Section~6) meets this need. Targeted training for graduate students and early-career
researchers---who are both forming citation habits and potentially developing software themselves---may yield disproportionate
long-term benefits. Senior researchers have an important complementary role:
explicit guidance on what citation standards to expect from their students and
postdocs can catalyze norm-setting at the group and department level.

For \textit{journal editors and publishers}, software citation policies should be strengthened. The
\href{https://www.chorusaccess.org/resources/software-citation-policies-index/}{CHORUS
software citation policies index} provides a reference point for existing
publisher policies;  research communities should encourage publishers without policies to adopt
them and those with weak ones to strengthen them \citep{Stall2023}.ArXiv could also perform lightweight software
citation checks at preprint submission: for instance, flagging papers that mention software names without corresponding reference-list entries, a
change that would reach the field at massive scale.

A complementary proposal is a volunteer network of software citation reviewers: researchers willing to be contacted by journal editors for assistance evaluating software \textit{citation} quality in submitted manuscripts. Such a network would lower the knowledge barrier that currently prevents many editors from enforcing citation standards. The success of the \textit{Journal of Open Source Software }\citep[JOSS;][]{JOSS...2017arXiv170702264S} and its open review process suggests that this could work well.

For \textit{funders and institutions}, the software impact metrics should explicitly be recognized in grant applications and
hiring/promotion processes. These metrics should not simply mirror article metrics like citation
counts but also software mentions in ADS, download and install statistics
(Zenodo, PyPI, conda-forge), and repository activity indicators (stars,
forks, upstream dependencies). The field lacks a standardized format for
presenting software impact evidence in CVs and grant applications; developing
such a format would lower the barrier for researchers to claim credit for
software contributions.

\subsection{Metrics and Recognition Beyond Citations}

Citation counts have well-known limitations as the sole metric of software
impact: they favor widely used tools over impactful niche ones, lag real
usage by months or years, and can be gamed. While formal citations do not need to be replaced, complementary metrics should also be considered. The key recommendation is that institutions and funders broaden
the evidence they accept. A researcher who can demonstrate that their code is
used in hundreds of papers---even if not always formally cited---has made a
contribution that deserves recognition. Separately,
the distinction between ``use,'' ``mention,'' ``build upon,'' and
``extend'' in the context of a software tool represents qualitatively different forms of
contributions, and that tracking these different relationships may ultimately
be more informative than aggregated citation counts alone.

%%
%% ─────────────────────────────────────────────────────────────────────────────
\section{Theme II: Best Practices, Conventions and Norms}
\label{sec:norms}
%% ─────────────────────────────────────────────────────────────────────────────
%%

Here we discuss the recommended best practices for software citation from the perspective of both researchers using software and software developers. While these are two distinct roles, both sets of best practices rely on each other to meet the needs discussed in Sections 1 and 2.

\subsection{Best Practices for Citing Software}

There are several key considerations for researchers when deciding how to cite software that they have used: (i) scope, (ii) dependencies, (iii) citation format, and (iv) versioning.

\subsubsection{Software Citation Scope}

One of the most persistent uncertainties for researchers is the question of
\textit{scope}: which software, out of the full stack used in a given
analysis, deserves formal citation? This question has no universal answer,
but the guiding principle is that \textit{citation should
function as a mechanism for credit and attribution}. A researcher deciding
which software to cite is making a judgment about whose intellectual and
technical labor was materially important to their results.

Software that served a generic infrastructure function---an operating system,
a compiler, a general-purpose text editor---typically does not satisfy this
criterion. Software that implemented a specific scientific algorithm, provided
the primary analysis framework, or enabled a key capability that the work
depended upon typically does. Standard community packages in a field (in
astronomy: NumPy, SciPy, Matplotlib, Astropy) are often expected to be
cited when used substantively, even when their role is ``supporting'' rather
than ``core.''

Community guidelines should explicitly state that
\textit{identifying what to cite is an active, considered process} and that the decision to cite (or
not) carries real consequences for the visibility and sustainability of
research software.

\subsubsection{Dependency Citation}

A closely related question concerns software dependencies. Most modern
scientific software is built on a stack of other packages, each representing
real intellectual contributions. How far down that dependency tree should
citation extend?

Dependency citation should not require a separate
framework---it belongs inside the same conceptual frame as direct software
citation \citep{Katz2014}. When a researcher decides which software to cite,
they are already making a choice about whose labor is visible. Dependencies
sit one layer removed from that decision, but the underlying logic is
identical: software that a scholarly work substantively relies upon represents
contributions that merit acknowledgment. The field therefore does not need
new norms for dependency citation so much as it needs existing citation norms
to explicitly encompass the question.

Dependencies are a category
of contribution that researchers ought to weigh as part of the active, considered process of choosing what to cite. Treating
dependency citation as an extension of established norms is the most tractable path toward consistent practice.

On the question of depth: the pragmatic recommendation is to cite
the highest-level packages used directly, plus any foundational dependencies
that played a distinctive or dominant role in the results. Citing every
transitive dependency is impractical and may not align with journal norms or
reviewer expectations. Importantly, \textit{dependency version information
does not need to appear in the formal citation}: version tracking for the
full dependency tree is more appropriately captured in a machine-readable
environment file (\texttt{requirements.txt}, \texttt{environment.yml},
\texttt{Pipfile.lock}, etc.), which is a matter of good software practice
rather than citation practice. Conflating the two creates unnecessary overhead
and can discourage dependency citation altogether.

\subsubsection{Citation Format and Machine Readability}

Software citations must appear in the formal reference list, not only in an
acknowledgment paragraph. The appropriate BibTeX entry types are
\texttt{@software} (preferred) or \texttt{@misc} where \texttt{@software}
is unsupported. Every entry should include a version number and a DOI
wherever available. Bare URLs are inadequate: they are not persistent, not
structured, and not machine-processable.

When new work both depends on existing software and includes a release of new software, authors are both researchers and developers, and all best practices should be followed appropriately. In addition to citing software when publishing an article, the code release of this type of work should include a machine-readable citation file such as \texttt{CITATION.cff} or
\texttt{codemeta.json} that describes all dependencies. This should not be considered an optional extra, but a standard component of
any publicly released scientific software package.

There is sometimes tension between \textit{preferred citations}
(as specified by developers in \texttt{CITATION.cff} files or documentation)
and \textit{conventional citation formats}. In principle, developers'
preferences should be followed. In practice, preferred citations sometimes
take inconvenient forms: pointing to a paper that is difficult to cite,
specifying an unusual format, or failing to include a DOI. In general, it is important that software citation tooling (especially the SCS) offer users the option to
output either the developer's preferred citation or a standardized
alternative, so that the absence of a well-formed preferred citation is not a
barrier to correct practice. The field should also develop clearer norms
about what constitutes a well-formed preferred citation, to guide developers
in crafting instructions that are both respected and practically followable.

\subsubsection{Versioning}

The same software can produce different results at different versions.
Software citations must therefore be version-specific: citing the exact
version (or commit hash, for development checkouts) used in the analysis.
This is separate from the dependency version question: while dependency
versions belong in environment records rather than reference lists, the
version of directly cited software must appear in the formal citation.
Versioned archival services such as Zenodo make version-specific citation
the default for each tagged release, reducing the burden on the researcher.

\subsection{Best Practices to Make Software Citable}

It is up to developers and/or maintainers of software to ensure that the artifacts and processes are in place so that it can be properly cited. This requires three main practices: assigning a persistent identifier, deciding what level of granularity that identifier applies to, and providing citation instructions.

The following sections describe each of these in more detail, but in short we recommend to:

\begin{enumerate}
    \item Archive the source code (and optionally installers/binaries) of each version of your software on Zenodo to get a DOI. Follow Semantic Versioning (SemVer 2.0.0) to decide when a change warrants a new version.
    \item Document how to cite your software in your README and in a CITATION.cff file.
\end{enumerate}

\subsubsection{Assigning a Persistent Identifier}

For a citation to function as a credit mechanism, it needs to point to something stable and trackable, not a repository URL that can move or be renamed. This is where persistent identifiers (PIDs) come into play. They are identifiers that are globally unique, resolvable, and can be used to reference your software unambiguously.

Amongst the various PID schemes, we believe Digital Object Identifiers (DOIs) are the strongest candidate to use for software due to their growing adoption (Figure 1) and supporting infrastructure for tracking citations. Registration agencies such as DataCite already track citations to DOIs, and journals already know how to require and parse them, which makes it ideal for citations to be surfaced back to you and receive credit for the reuse of your software. Zenodo, a generalist repository supported by the European Organization for Nuclear Research (CERN), is the most commonly used to obtain a DOI for software, accounting for over 90\% of software DOIs.

\begin{figure}
    \centering
    \includegraphics[width=1\linewidth]{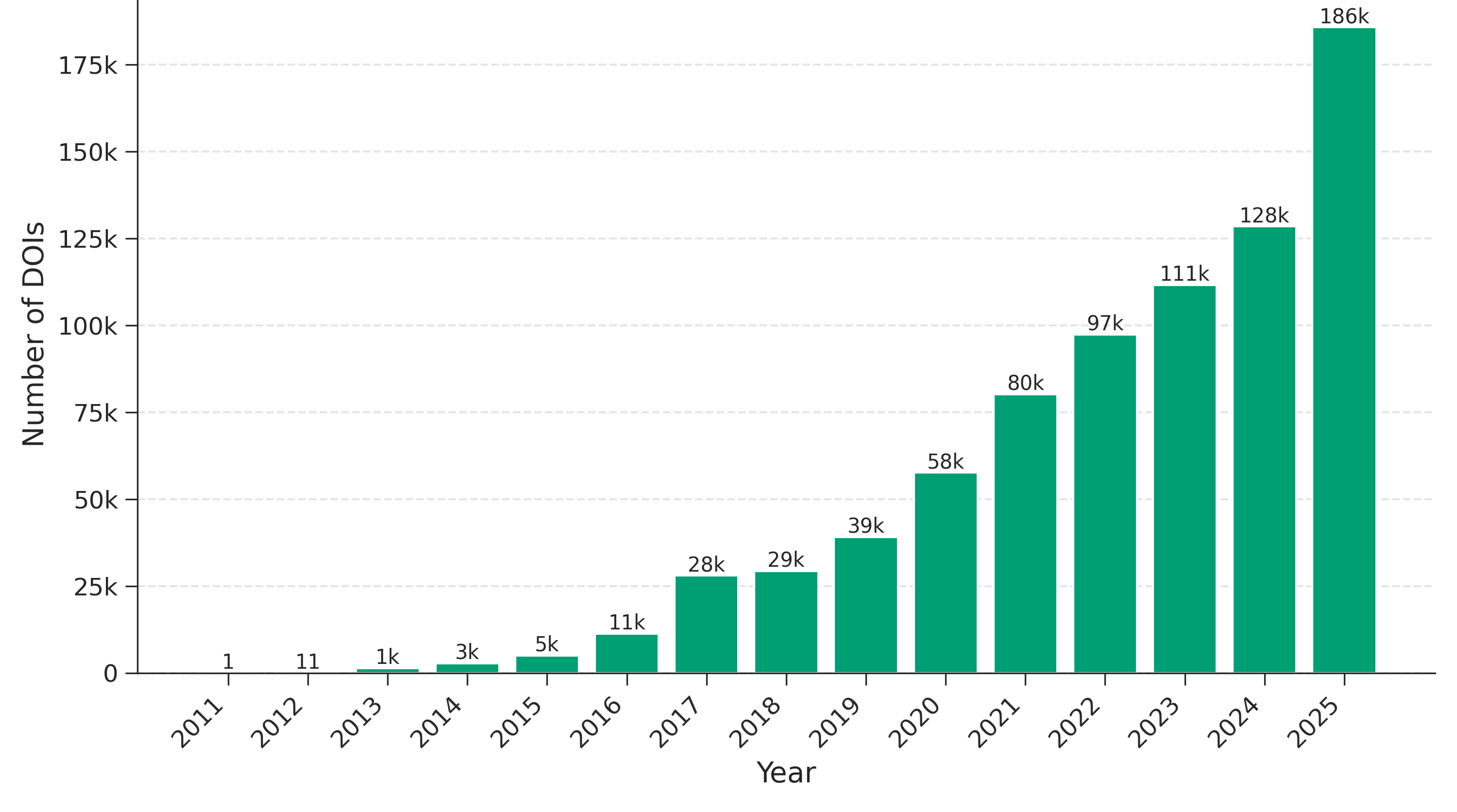}
    \caption{Number of software DOIs generated through DataCite by year.}
    \label{fig:doiYear}
\end{figure}

\subsubsection{Choosing Levels of Granularity}

One may wonder what should be assigned a PID and made citable. Should the PID be assigned to the source code, compiled installer, or something else? Again, there is no established standard here. It is up to the developers of the software to decide what level of granularity in their software needs to be distinguished unambiguously and made citable based on the reuse cases they anticipate. A common practice we have observed is each GitHub release being archived on Zenodo and assigned a DOI.

The FAIR Principles for Research Software (FAIR4RS Principles) prescribe that ``Different versions of the software are assigned distinct identifiers'' (F1.2). This is because each version carries different features but also different authors, as mentioned before, and therefore making each version citable distinctly is essential for properly crediting contributors of software. Accordingly, we recommend that, at the very least, the source code for each version of your software should be assigned a PID. \href{https://semver.org/}{Semantic Versioning} (SemVer 2.0.0) is a widely adopted convention for versioning software that we recommend.

\subsubsection{Providing Citation Instructions}

Even with a PID in place, users of your software still need to know about it. We recommend that you provide citation instructions in at least two ways that together provide both human-friendly and machine-actionable citation instructions. 

First, we recommend that you provide citation instructions in your README. You can include them in a ``How to Cite'' or “Citation” section, for instance. If you obtained a DOI through Zenodo, you can also just refer to the ``Citing'' section of Zenodo, which provides citation guides in several styles (APA, Chicago, etc.) and export options in several formats (JSON, BibTeX, etc.). 

Second, we recommend including a CITATION.cff file \citep[][]{citation-cff} at the root level of your source code. A CITATION.cff file is a structured, YAML-based Citation File Format (CFF) that captures the citation information (author names, title, version, DOI, license, release date, and more) in a machine-actionable form. GitHub natively recognizes CITATION.cff files: when one is included in your repository, GitHub displays a ``Cite this repository'' widget that auto-generates citation text, which can lower the barrier for reusers to cite correctly and consistently.

Note that there exists a GitHub-Zenodo integration that lets authors mint a DOI automatically on every tagged release on GitHub. However, that integration does not provide the DOI in advance, so you cannot update your CITATION file with the latest DOI before those files get archived on Zenodo. Therefore, we recommend instead tools including  \citep[][]{codefair_portillo_2026_19261487} or the TARDIS reduce pipeline (Section~6), which are open source platforms developed to address this gap. It makes it easy to create GitHub releases and archive them simultaneously on Zenodo while automatically updating your CITATION.cff with the latest DOI before the Zenodo archive is released.

%%
%% ─────────────────────────────────────────────────────────────────────────────
\section{Theme III: Infrastructure and Workflow}
\label{sec:infrastructure}
%% ─────────────────────────────────────────────────────────────────────────────
%%

One can imagine a world in which software citation is so thoroughly and smoothly
embedded in the research workflow that it takes conscious effort from a
researcher to \textit{fail} to cite software correctly. In this case, software citation infrastructure would likely be integrated at all stages
of the research lifecycle: from APIs and data formats that bake in the relevant
citation metadata, to bibliographic databases that include software on equal
footing with journal articles, to authoring tools that intelligently combine all
this information to insert citations into articles, to promotion and tenure
standards that implicitly expect software to be a common form of research
product. In this world, the software citation workflow is indistinguishable from
the research workflow in general. It will take a long time to make this world a
reality.

\subsection{The ``Golden Path''}

How do we get from the current, real world to this idealized future? A useful concept to consider is the notion of a \textit{golden path} --- a
workflow that makes correct software citation the path of least resistance for a
researcher finalizing a manuscript. The golden path is insufficient to achieve
software citation nirvana: if you only think about software citation at the last
stages of a project, there are inevitably going to be frictions present that
could have been avoided if you had thought about it continuously from the start.
But the existence and visibility of such a path would represent a significant
step forward compared to the status quo. Ideally, the golden path would be
straightforward enough that researchers who are willing to cite correctly will
reliably follow it; if it requires context-switching between tools or manual
effort at each step, even motivated researchers will default to shortcuts.

The current landscape lacks a clear golden path. A researcher who wants to cite
software correctly must gather information about the software they have used,
make use of multiple tools, navigate
between their manuscript editor and external websites, and manually format the
resulting references. Every step that requires context-switching is an
opportunity for the workflow to break down.

There are three complementary strategies for working towards the
creation of a single golden path. First, \textit{aggregation}: bringing citation
lookup into a single, comprehensive tool --- rather than requiring researchers
to know which registry to consult for which package. Second,
\textit{integration}: embedding citation lookup directly into the authoring
environment (manuscript editors such as Overleaf, or browser extensions for
web-based tools). Finally, \textit{automation}: prepopulating citation
suggestions based on package imports or dependency files detected in a code
repository.

\subsection{Breadth and Diversity of The Existing Ecosystem}

Considering the many forms of research infrastructure, a complete review of the existing ecosystem for software citations is effectively
impossible. That being said, there are some pieces of research infrastructure
that are especially relevant to the software citation golden path.

Perhaps first to come to mind are the \textit{repositories}: services that allow
software to be ``deposited'' in some fashion that ensures its long-term
availability. Currently, the highest-profile of these is \textbf{Zenodo} \citep[][]{zenodo}, the
CERN-supported digital repository. Among other things it can be used to
archive software releases and mint DOIs, with a GitHub integration that can
make the deposition process fairly automatic. Zenodo's DOI versioning model (a
``concept DOI'' identifying the software as a whole; ``version DOIs''
identifying individual releases) is the recommended approach for versioned
software citation.

Many modern programming languages are associated with package repositories that
serve similar purposes: NPM for JavaScript, Crates.io for Rust, etc. These
repositories, however, are often not designed with scholarly uses in mind, which
constrains their ability to support software citation. In particular, DOI
generation, formalized author lists, and deposition of metadata to services like
\textbf{CrossRef} \citep[][]{hendricks_crossref_2020} may be lacking. The \textbf{Comprehensive R Archive Network} \citep[\textbf{CRAN};][]{CRAN} is a good example of a mature package repository and distribution network, and \textbf{Bioconductor} \citep[][]{gentleman_bioconductor_2004,Huber...bioconductor...2015} is an R package ecosystem
for bioinformatics that \textit{does} provide a
mature citation infrastructure.

While repositories store software, \textit{registries} are services
that issue unique, persistent identifiers (PIDs) that can be attached to
software or other resources. Any effective software citation ecosystem must be
built on a robust PID foundation. The \textbf{Digital Object Identifier} \citep[\textbf{DOI};][]{DOIHandbook}
system is the canonical example of such a system. The \textbf{Research Resource
Identifier} \citep[RRID;][]{RRID} system is generally accepted in the life sciences but rarely
used in other disciplines. The \textbf{Astrophysics Source Code Library} \citep[\textbf{ASCL;}][]{Allen2016ascl}
plays a similar role in the astronomy community, with limited support for acting
as a software repository. \textbf{Software Heritage} \citep[][]{software...heritage}, on the other hand,
aggressively embraces both roles: it collects and preserves software in source
code form, and also computes and publicizes unique Software Hash Identifiers \citep[SWHIDs;][]{SWHIDs...2023arXiv231010295D}. SWHIDs are quite flexible and can identify software releases
(versions), commits, directories, files, or even source code fragments. In a
software-only alternate world, SWHIDs would be good identifiers to use for
software citation, but in our current world where DOIs have become the standard
for papers and data, it is not clear if there is room and appetite in the
publishing and indexing community to support SWHIDs.

Registries are, in turn, indexed by \textit{bibliographic databases} that offer
services such as discovery, citation counting, and analytics. In astronomy, the
\textbf{Astrophysics Data System / Science Explorer} \citep[\textbf{ADS/SciX};][]{ADS} is the dominant
system, with high-quality support for software citations developed over a number
of years. Other systems like \textbf{PubMed} \citep[][]{NCBIresource_pubmed}, \textbf{Google Scholar} \citep[][]{GoogleScholar_contributions},
\textbf{Semantic Scholar} \citep[][]{semantic_scholar}, or \textbf{Scopus} \citep[][]{scopus} are generally targeted at journal
articles and do not integrate software citations as well. The \textbf{Asclepias} \citep[][]{asclepias...2020ASPC..522..711M}
project, developed in conjunction by the American Astronomical Society (AAS),
Zenodo, and ADS/SciX, provides a database of links between citable software and
journal articles.

We define \textit{software citation portals} as services that help users
discover how to cite specific pieces of software. The search functions of
repositories, registries, and bibliographic databases can support this
functionality, but often are not designed to guide users through specific
workflows. This is the category that \textbf{Software Citation Station \citep[SCS;][]{Wagg2024}}
fits into. So does \textbf{CiteAs} (\href{https://citeas.org/}{https://citeas.org/}), a more generic service built on top of
\textbf{OpenAlex} \citep[][]{OpenAlex...2022arXiv220501833P}. It is also worth pointing out that \textbf{generic web search
engines} and \textbf{generic LLM chat bots} fit this definition of a portal,
mainly to emphasize that a purpose-built portal had better provide demonstrably
superior functionality if it hopes to see serious usage. 

\textit{Scholarly journals} have an important role to play in software citation:
you might have a fistful of software DOIs ready to cite, but if your preferred
journal only allows articles to cite other articles, they won't do you any good.
Historically, some journals have also served as de-facto software repositories,
allowing source code to be attached to published articles. Modern usage
recognizes that it is best to deposit software in purpose-built repositories.
The \textbf{Journal of Open Source Software} \citep[JOSS;][]{JOSS...2017arXiv170702264S} merits special mention as a
``software journal'' that supports the peer review and publication of software
in a way that is legible to the traditional, article-centered citation
ecosystem.

\textit{Authoring tools} can be viewed as yet another element of software
citation infrastructure, especially as authoring occurs more and more on managed
platforms like \textbf{Overleaf} (\href{http://overleaf.com}{http://overleaf.com}), \textbf{Curvenote} \citep[][]{curvenote...cockett2024}, \textbf{PubPub} (\href{https://www.pubpub.org/}{https://www.pubpub.org/}), or even
\textbf{Google Docs}, rather than traditional, locally-installed word
processors. The golden path of software citation should bring citation
assistance into the existing writing workflow via plugins that can automatically
suggest software citations based on detected package names in code blocks.
Integration with \textit{reference managers} such as \textbf{Zotero} \href{https://www.zotero.org/}{https://www.zotero.org/} and 
\textbf{Mendeley} \href{https://www.mendeley.com/}{https://www.mendeley.com/} would similarly reduce friction.

Finally, \textit{standards} should be recognized as essential components of the
infrastructure to support software citation. Particularly relevant are data
standards like the \textbf{Journal Article Tag Suite} \citep[\textbf{JATS};][]{JATS2024}, which most
scholarly journals use ``under the hood'' to represent articles, including their
citations; the \textbf{Citation File Format} \citep[CFF;][]{citation-cff} intended to link source code
repositories with external citeable objects; and the \textbf{CodeMeta} \citep{Jones2017codemeta}
vocabulary for capturing metadata about source code, including citation
information.

\subsection{Responsibility for Infrastructure}

A recurring practical question is who bears the responsibility for maintaining,
funding, and developing software citation infrastructure. There is no one definitive answer, but several principles that should be followed.
Infrastructure tools should, where possible, be built on open standards and
open data to reduce single-point-of-failure dependencies. Funding
responsibility should be distributed across institutions (universities,
national labs), research funders (NSF, NASA, ESA, private foundations), and
publishers who benefit commercially from the research that software enables.
The sustainability of volunteer-maintained
tools is a concern, so key components of software citation
infrastructure should be placed on stable, institutionally supported footing.

%%
%% ─────────────────────────────────────────────────────────────────────────────
\section{Theme IV: Citing AI-Assisted Research and Emerging Challenges}
\label{sec:ai}
%% ─────────────────────────────────────────────────────────────────────────────
%%
The growing integration of large language models (LLMs) and AI-assisted tools into research workflows raises questions for citation practice that conventional software citation mechanisms do not fully address. LLM outputs are stochastic: the same prompt to the same model can yield different results across runs and a measurable output variable persists even under normally deterministic settings \cite{Atil2025nondeterminism}. Commercial models served through APIs can also change under the hood, with documented behavioral drift between successive versions of the ``same'' model within months and deprecated models can be withdrawn entirely \cite{Chen2023drift, anthropic2025deprecation}. Finally, systematic audits of major foundation model developers found widespread deficiencies in transparency regarding upstream resources, including training data, data labor, and compute. The average transparency score was just 37/100 in the initial October 2023 Foundation Model Transparency Index and increased only to 58/100 in the 2024 follow-up, with persistent opacity remaining in areas such as data labor and data access \cite{Bommasani2023fmti, Bommasani2024fmti}. A citation that names a model and a version number therefore pins down less about what actually influenced a paper than an equivalent citation to a versioned, archived software release. 

Scholarly publishing is gradually converging on several common principles: generally, AI systems cannot be listed as authors because authorship entails responsibility and accountability that a model cannot assume. Journals increasingly require disclosure of substantive AI use, including the tools used and how they contributed, while making clear that human authors remain fully responsible for all content, however it was generated (\cite{ICMJE2024, COPE2023, NatureEditorial2023, Thorp2023}). However, current real-world practice is not fully aligned with the policy. A distributional estimate of LLM-modified text indicates that by late 2024, roughly 22\% of computer science abstracts showed signs of substantial LLM modification, compared with under 10\% in mathematics and the Nature portfolio journals \cite{Liang2025quantifying}, and that between 6.5\% and 16.9\% of reviews submitted to major machine learning conferences had been substantially modified by LLMs \cite{Liang2024monitoring}. More recently, repositories such as arXiv have strengthened enforcement against unchecked AI-generated content, including imposing submission bans where manuscripts contain clear evidence that authors did not verify machine-generated material, such as hallucinated references. \cite{arXivModeration}.

Conventional citation mechanisms, such as citing the model as software with a version number and DOI, can capture some relevant information, but may miss the specifics on how the model was used, what role it played in generating results and what is known about its training data. Meanwhile, disclosure requirements across major publishers and venues converge on substance while diverging on placement. The ICMJE directs authors to describe AI use in both the cover letter and the manuscript itself, with research applications documented in the Methods and writing assistance noted in the Acknowledgments~\cite{ICMJE2024}, and COPE similarly locates disclosure in the Methods or another appropriate section~\cite{COPE2023}. Springer Nature requires a Methods statement or, where a paper has no Methods section, a suitable alternative~\cite{NatureEditorial2023}, whereas Science asks for disclosure in the cover letter and Acknowledgments~\cite{Thorp2023}. Elsevier departs from both by mandating a dedicated AI declaration positioned immediately before the reference list~\cite{Elsevier2023genai}. ACM scales the requirement to the extent of AI involvement, accepting Acknowledgments and in-text disclosure as appropriate~\cite{ACMauthorship}, and IEEE asks that the Acknowledgments name the system used and identify the affected portions of the manuscript~\cite{IEEEai}. arXiv, rather than specifying a disclosure location, enforces author responsibility at the point of submission and moderation~\cite{arXivModeration}. Across all of these, AI systems are barred from authorship, human authors retain full responsibility for published content, and minor copy editing such as grammar and spelling correction is generally exempt from disclosure altogether. Meanwhile, existing software citation also falls short: CITATION.cff and CodeMeta lack standardized fields for LLM-specific metadata such as model state, sampling parameters, and training-data provenance while CRediT contributor taxonomy is designed around human contributors \cite{CRediT}. Therefore, a separate, voluntary disclosure mechanism is more appropriate for this purpose, complementing rather than replacing standard citation practice.

\subsection{The Model Influence Statement}
\label{ai:the_model_influence_statement}

Here we describe a framework for a \textit{Model Influence Statement}: a structured, voluntary disclosure accompanying a paper that describes the role of AI models in the research and writing process \cite{Shan_Model_Influence_Statement_2026}. The statement opens with a responsibility declaration where the listed authors remain solely responsible for the work and for verifying that any AI-generated code or text performs as intended, and proceeds through a structured sequence: whether models were used at all, with a signed attestation even in the negative case; a repeatable metadata block for each disclosed model covering model name, developer, version, license, base model, links to documentation and training-data descriptions, persistent identifiers and estimated CO$_2$ emissions \cite{Lacoste2019co2}; a training-data disclosure indicating whether the model was trained on open-source code or restricted data; the roles the model played, expressed with reference to the CRediT taxonomy but open to free text; optional disclosure of critical prompts; ethical considerations and a signed certification.

Two design decisions distinguish the framework. First, ``unknown'' is a first-class answer throughout. Given the documented opacity of commercial model development \cite{Bommasani2024fmti}, a disclosure format that demands information authors cannot obtain could lead to non-compliance or fabrication. Putting ``unknown'' can make honest disclosure achievable while still recording the limits of what is knowable. Second, the framework's instructions explicitly direct reviewers to value the transparency a statement reflects without treating its presence, absence, or level of detail as a standalone signal of research quality. A prototype web-based generator for these statements is available at \url{https://model-influence-statement-generator.netlify.app/}. It is important that this remain voluntary in the near term, allowing community norms to develop organically before standardization. While cross-publisher efforts are now in consultation, the design of Model Influence Statement is intended to map cleanly onto whatever consensus format emerges.

\subsection{AI as a Citation Tool}
\label{ai:ai_as_a_citation_tool}

LLMs can also assist with software citation rather than merely serving as an object of it. The gap such assistance could close has been researched: studies of software in the literature consistently find that informal mentions vastly outnumber formal, machine-actionable citations \cite{Howison2016seeing, Du2021softcite, Schindler2023}, and large-scale extraction efforts have recovered significant mentions invisible to citation counts \cite{Istrate2022czi}. 

There is strong potential for AI-assisted tools to help researchers inventory the software they used, locate recommended citations, and generate formatted BibTeX entries---automating significant
parts of the citation workflow. Tools such as Citey (Section~6) incorporate this approach. As AI assistance in citation becomes more prevalent, it will be important to develop standards for disclosing when a citation was generated by an AI tool rather than manually verified by the researcher. A provenance flag that distinguishes machine-suggested from human-verified references would let readers, reviewers and indexers weight references accordingly, and would extend Theme III's infrastructure (Section~4) discussion to the citation-production process. A useful recursion can be found here: an AI-assisted citation workflow is itself a model user, and the Model Influence Statement of Section 5.1 provides exactly the disclosure channel through which it can be reported.

%%
%% ─────────────────────────────────────────────────────────────────────────────
\section{Open Questions and Future Directions}
\label{sec:future}
%% ─────────────────────────────────────────────────────────────────────────────
%%

There are several questions that remain unresolved and should
be priorities for future community attention.

\paragraph{The role of reviewers}
Journal peer reviewers represent a natural checkpoint for software citation quality,
but the field lacks norms, training, or incentive structures for reviewers to
play this role. There are two distinct reviewer functions:
evaluating whether software \textit{citation statements} are complete and
correctly formatted, and evaluating whether the \textit{software itself}
meets quality standards: a function closer to JOSS's model. Both are
valuable; neither is currently standard practice. We recommend
developing explicit reviewer guidance and exploring a volunteer reviewer pool that journals could consult for software citation feedback.

\paragraph{Dependency citation: norms still forming}
Despite the framework in Section~3, significant variation
remains across disciplines in how far down the dependency tree citation
extends. Practices in astronomy differ from those in biology, where
Bioconductor has established stronger norms. Longitudinal tracking of
citation patterns via ADS could help the community understand how norms are
evolving and where systematic gaps persist.

\paragraph{Alignment with FAIR software}
The \href{https://www.nature.com/articles/s41597-022-01710-x}{FAIR Principles
for Research Software} \citep[FAIR4RS;][]{WilkinsonFAIR,Katz...FAIR...2021arXiv210110883K,ChuHong2022fair4rs,Barker2022} provide a complementary
framework for making software Findable, Accessible, Interoperable, and
Reusable. Software that meets FAIR4RS criteria is, in most cases, also easy
to cite. Software citation guidance should be explicitly connected to FAIR4RS, and exploring whether citability could be incorporated
as a measurable dimension of FAIR compliance would be helpful.

\paragraph{Identifier redundancy}
Multiple independent systems may independently assign identifiers
to the same software: e.g., general Zenodo; astrophysics-specific, such as ASCL and ADS; and others in other disciplines, such as SWMath IDs in math and RRIDs in life sciences. This creates fragmentation and confusion. Better coordination between registries and repositories, and clearer guidance to users on which identifier to prefer and when, would reduce this problem.

\paragraph{Commercial and non-standard software}
Widely used tools such as Excel and Overleaf occupy an ambiguous space:
they are used in scientific research but are not archival scientific software in the same sense as a simulation code or analysis pipeline. There is no recommended standard on whether and how such tools should be cited, so this should be decided via
community discussion and field-specific guidance. Adding commercial software
to registries such as the SCS via RRID was discussed as one possible approach to make it easier for researchers to cite such software if they prefer to.

\paragraph{The ArXiv opportunity.}
ArXiv receives essentially all preprints in astronomy. A lightweight software
citation check at preprint submission, flagging papers that mention software
names without corresponding reference entries, could reach the field at
a scale no other single intervention could match. A collaboration with ArXiv to pilot such a check would be useful.

\paragraph{Impact of suggested solutions.}
The interventions described in this paper are hypotheses about what will
work. Systematic tracking of whether and how citation practice changes in
response to new tools, guidelines, and policies is essential for the
community to know what is actually working. Evaluation should be directly built into the deployment of each new tool or policy.

\paragraph{Citing AI models that no longer exist.}
Commercial AI models are routinely deprecated and withdrawn, normally with no archival path analogous to Zenodo or Software Heritage. While a cited model ceases to exist, a citation to it can become unresolved in a way that persistent-identifier ecosystem cannot currently repair. Whether and how closed-weight model states should be archived, by whom and under what identifier scheme is an open question that software citation community shares with the broader AI governance community.

%%
%% ─────────────────────────────────────────────────────────────────────────────
\section{Conclusions}
\label{sec:conclusions}
%% ─────────────────────────────────────────────────────────────────────────────
%%

Software citation in (computational) research s is broken, but it is not
irreparably so. The technical components needed for good practice---persistent
identifiers, structured metadata formats, aggregation tools, and automated
workflows---mostly exist, with AI-mediated contribution a clear exception. What is missing is the integration, visibility, and
cultural weight that would make correct citation the path of least resistance.

Here we describe three overarching conclusions. First,
progress requires simultaneous action on technical and cultural fronts:
tooling improvements without cultural change will not move the needle, and
cultural advocacy without better tooling will not be sustained. Second,
journal editors and publishers represent the single highest-leverage point
for accelerating cultural change; engaging them directly, through collective
letters, reviewer training, and formal policy adoption, should be a
community priority. Third, the path to correct software citation must be
brought into the researcher's existing workflow. A golden path that requires
context-switching between multiple external tools, or that is invisible to
researchers who do not already know to look for it, will not achieve the
scale of adoption the field needs. These conclusions can now often be acted on in a research environment that is being reshaped by AI-assisted workflows. That reshaping cuts both ways: model-assisted research strains citation practice in ways existing mechanisms were never designed to handle, while the same tools offer the most plausible route yet to citation assistance that is actually frictionless. The community's window to establish norms is open while those norms are still forming, and it will not stay open indefinitely.

The tools and recommendations described in this paper are a starting point.
We invite the broader community---researchers, developers, publishers,
funders, and tool builders---to engage with these outputs, contribute to
their development, and help build the norms and infrastructure that software
citation deserves. Software is a first-class scientific contribution; its
citation practices should reflect that.

\section*{Data Availability}

No data is shared as part of this article.

\section*{Code Availability}

No custom code was used as part of this article. All referenced code is publicly available in the references links.

%%
%% ─────────────────────────────────────────────────────────────────────────────
\section*{Funding}
%% ─────────────────────────────────────────────────────────────────────────────
%%

The authors acknowledge support for the 2026 Software Citation Workshop from the NASA TWSC program under award number 316634-00001. 
P.R.V. and F.S.B. acknowledge support from NASA HPOSS grant 80NSSC25K7555 under award number 316592-0000. P.R.V. was supported by a Postgraduate Scholarship - Doctoral (PGS-D) Award [PGS D - 589175 - 2024] from the Natural Sciences and Engineering Research Council of Canada (NSERC). F.S.B acknowledges support from NSF 22-624 Astronomy and Astrophysics Research Grants under award number 2606407. A.Z. was supported by NSF Mid-Scale Research Infrastructure-2 grant \#2153201 (PI: Mark Devlin, University of Pennsylvania) for the Advanced Simons Observatory; Simons Foundation Award \#457687. L.M.D. acknowledges financial support from the Alexander von Humboldt Foundation. A.A. acknowledges support from Heidelberg Institute for Theoretical Studies, University of Maryland College Park, and Michigan Technological University.

\bibliographystyle{apsrev4-1}
\bibliography{software_citation_whitepaper}

%%
%% ─────────────────────────────────────────────────────────────────────────────
\begin{appendix}
%% ─────────────────────────────────────────────────────────────────────────────
%%

%%
%% ─────────────────────────────────────────────────────────────────────────────
\section{Resources developed during the workshop}
\label{sec:outcomes}
%% ─────────────────────────────────────────────────────────────────────────────
%%

Here we describe several resources, each
targeting a specific gap identified in this work, that were developed during the Software Citation Workshop.

\paragraph{Software Citation Primer.}
A short, practical, two-part guide: a seven-step workflow for researchers
citing software in manuscripts, and a three-step checklist for developers
making their software citable. Designed to be read in a single sitting and
to be immediately actionable, the Primer addresses the knowledge deficit
identified across all stakeholder groups. Available at \citet{Broekgaarden:2026-software-primer}.

\paragraph{Improved version of the Software Citation Station.}
We improved the Software Citation Station \citep{Wagg2024}, including:
 a programmatic API
enabling downstream tools (Citey, Overleaf plugins) to query the Software Citation Station data;
output options supporting both preferred and standardized citation formats;
and journal-style modes respecting field-specific formatting conventions.

\paragraph{Citey: in-flow software citation.}
A browser extension that enables researchers to generate formatted software
citations---BibTeX entries, acknowledgment text, and Zotero-compatible
records---without leaving their writing environment. The user highlights a
software name, clicks one button, and receives export-ready citations.
Citey uses a pre-built, curated JSON database for instant response (no
network call on the hot path), with live fallback lookups for packages not
in the local database. A contribution prompt creates a growth flywheel:
researchers who encounter a missing package are routed to a page that helps
them make their own software citable, building the open citable-software
corpus over time. Developed by Jonathan Starr, Lucas de Sa, and Daina
Bouquin. Code: \url{https://github.com/jring-o/citey} \citep[][]{starr_citey_2026}. Web:
\url{https://citey.scios.tech}. 

\paragraph{Model Influence Statement framework.}
A structured, voluntary disclosure framework for AI-assisted research, accompanied by a prototype web-based generator. Authors record a responsibility declaration, per-model metadata (version, license, base model, training-data provenance, persistent identifier), the roles each model played, optional critical prompts, and ethical considerations. Two design choices make honest disclosure achievable: ``unknown'' is a valid answer throughout, reflecting the real opacity of commercial models; and reviewers are directed not to treat the presence or detail of a statement as a quality signal. The generator exports both a full statement and a one-paragraph acknowledgment summary. Described in detail in Section 5. Developed by \citet{Shan_Model_Influence_Statement_2026}. Generator: \url{https://model-influence-statement-generator.netlify.app}.

\paragraph{TARDIS automated release pipeline.}
A template GitHub Actions workflow for automating software releases with full
citation metadata maintenance: Zenodo archival and DOI minting, CITATION.cff
updates, contributor attribution via mailmap and ORCID, and bot-driven prompts
for contributor information at commit time. The pipeline reduces the per-release
overhead of maintaining citation metadata to near zero and serves as a
reference implementation of best practice. Available at
\url{https://github.com/tardis-sn/release-pipeline-template}.

\end{appendix}

\end{document}